\documentclass[aps,prb,superscriptaddress,reprint,showpacs,floatfix]{revtex4-2}
\usepackage{graphicx}
\usepackage{dcolumn}
\usepackage{bm}
\usepackage{xcolor}

\usepackage[T1]{fontenc}
\usepackage{relsize}
\usepackage{amsmath}
\usepackage{amsfonts}
\usepackage{amssymb}
\usepackage{mathtools}
\usepackage{braket}
\usepackage{gensymb}
\usepackage{url}
\usepackage[colorlinks = true,
            linkcolor = blue,
            urlcolor  = blue,
            citecolor = blue,
            anchorcolor = blue]{hyperref}
\usepackage{footmisc}
\usepackage[utf8]{inputenc}

\begin{document}

\preprint{APS/123-QED}

\title{Extended XY model for spinor polariton simulators}

\author{A. Kudlis}
\affiliation{Science Institute, University of Iceland, Dunhagi 3, IS-107 Reykjavik, Iceland}
\author{D. Novokreschenov}
\affiliation{Russian Quantum Center, Skolkovo, Moscow, 121205, Russia}
\affiliation{Abrikosov Center for Theoretical Physics, MIPT, Dolgoprudnyi, Moscow Region 141701, Russia}
\author{I. A. Shelykh}
\affiliation{Science Institute, University of Iceland, Dunhagi 3, IS-107 Reykjavik, Iceland}

\date{\today}

\begin{abstract}

The classic lattice XY model is one of the universal models of statistical mechanics appearing in a broad variety of optical and condensed matter systems. One of its possible realizations is a system of tunnel-coupled spinor polariton condensates, where phases of individual condensates play a role of the two- dimensional spins. We show that the account of the polarization degree of freedom of cavity polaritons adds a new twist to the problem, modifying in particular the structure of the ground state. We formulate the corresponding classical spin Hamiltonian, which couples phase and polarization dynamics, and consider several particular geometries, demonstrating the principal differences between the scalar and spinor cases. Possible analog of spin Meissner effect for coupled condensates is discussed. 

\end{abstract}

\maketitle

\textit{Introduction.} Lattice models are ubiquitous in modern physics covering a wide range of its fields from classical material science to cold atomic gases \cite{Bloch2008}, quantum magnetism \cite{Balents2010} and high-temperature superconductivity \cite{PatrickLee2006}. Although looking simple, they nevertheless demonstrate a large variety of emergent
phases and associated phase transitions \cite{Moreo1990,Giamarchi1991,Mizusaki2006,Samlodia2024}, including but not limited to superfluids and Mott insulators 
\cite{Greiner2002}, Tonks-Girardeau bosonic
gas \cite{Kinoshita2004,Paredes2004}, charge \cite{Bedcca2000} and spin \cite{Chang2010} density waves, singlet and triplet superconductivity \cite{Lichtenstein2000,Eichenberger2007},  stripe \cite{Xu2022} and supersolid \cite{Iskin2015} phases.

The lattice XY model is one of the cornerstone models of statistical mechanics \cite{Kosterlitz1974}. It describes a universal behavior of a 2D system, classical or quantum, with continuous symmetry \cite{Nelson1977,Frohlich1988,Ladewig2020}, and predicts, in particular, the emergence of the corresponding Berezinsky-Kosterlitz-Thauless (BKT) phase transition \cite{Kosterlitz1974,Gupta1988,Ding1992,Hu2011}. XY-type models naturally appear in such fields as low-dimensional magnetism \cite{Mattis1979,Duxbury1981}, 2D superfluidity \cite{Nelson1977,Mon1980}, ultra-cold bosonic quantum gases in optical lattices \cite{Struck2021,Sbierski2024},
superconducting systems \cite{Gingras1996,Xu1998,Babaev1999,Franz2006}, liquid crystals \cite{Shahbazi2006,Drouin2022} and even biophysics \cite{OHern1999}. In addition to fundamental physical interest, it was recently shown that XY models can be directly applicable in information processing algorithms, in particular, in problems related to machine learning \cite{Stroev2021}.

Coupled exciton-polaritons condensates have recently been proposed to represent a versatile platform for the realization of a classical XY model with tunable properties \cite{Lagoudakis2017,Berloff2017,Suchomel2018,Kalinin2020,Peng2024}. In this case, each condensate is characterized by its well-defined phase, equivalent to a two-dimensional spin, and the spin-spin interaction is provided by tunnel coupling of the condensates. The coupling constant depends on the geometry of the system, in particular the distance between the condensates \cite{Ohadi2016,Kalinin2018,Kalinin2018_2,Topfer2020,Cherotchenko2021}, and can thus be easily tuned in experiment \cite{Alyatkin2020}. 
\begin{figure}[t]
\includegraphics[width=1\columnwidth]{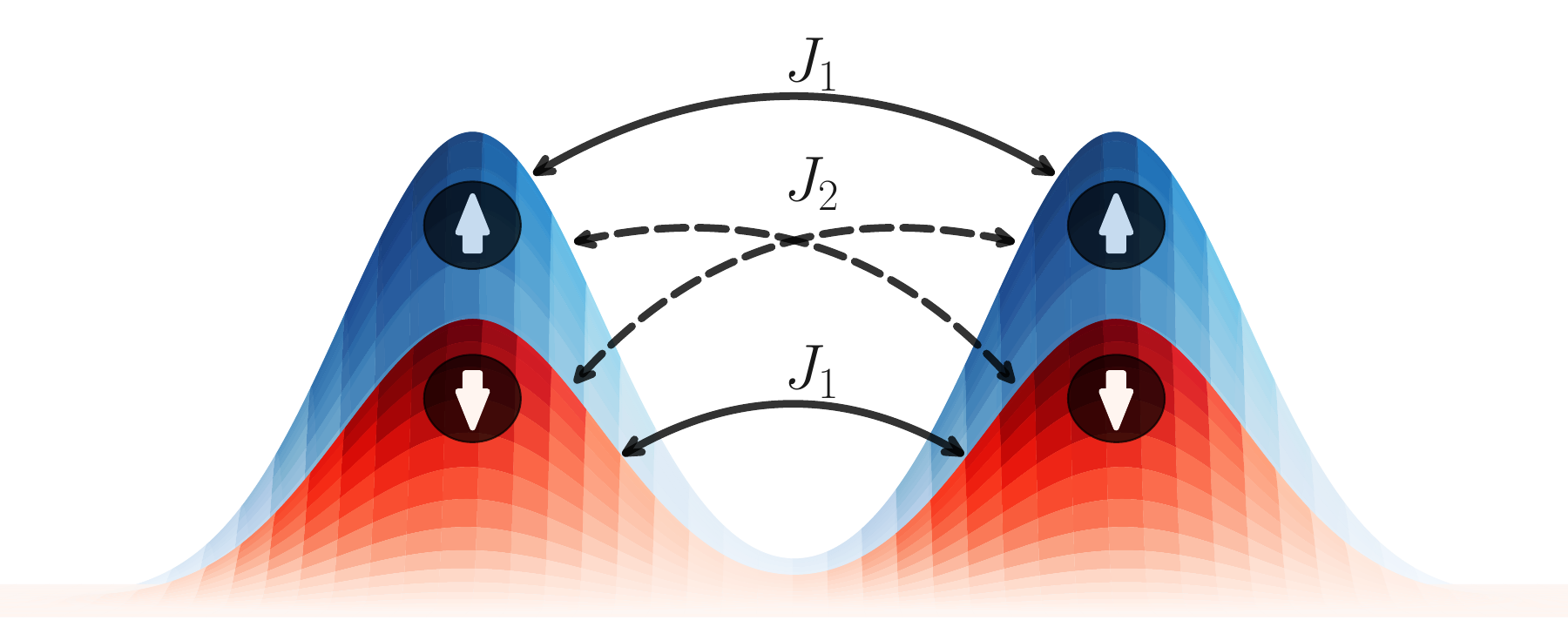}
\caption{Schematic illustration of the two distinct coupling mechanisms considered in the extended XY model. Solid lines represent spin-conserving tunneling ($H_1$ in~\eqref{eq:classical_ham}), characterized by coupling constant $J_1$, where particles maintain their polarization when hopping between sites. Dashed lines denote polarization-flip tunneling ($H_2$ in~\eqref{eq:classical_ham}), characterized by coupling constant $J_2$ induced by TE-TM splitting, which causes particles to change their polarization state during inter-site transfer. Blue and red bell-shaped figures correspond to right and left circular polarized components of two coupled condensates. 
\label{fig:1}}
\end{figure}

The study of XY polariton simulators was almost exclusively focused on the case of scalar polaritons, where the polarization degree of freedom was neglected. However, the latter is an important property that governs various aspects of polariton physics \cite{ShelykhReview}.
Similarly to photons, polaritons have two possible spin projections on the structure growth axis corresponding to the two opposite circular polarizations, which can be mixed by effective magnetic fields of various origin, such as in-plane directed magnetic field provided by TE-TM splitting and external magnetic field directed perpendicular to the structure interface. Importantly, polariton-polariton interactions are spin-dependent as well.

Indeed, they stem from the interactions of their excitonic components, which is dominated by the exchange term \cite{Ciuti1998}.
This leads to the fact that polaritons of the same circular polarization interact orders of magnitude stronger than polaritons with opposite circular polarizations \cite{Glazov2009}. As polariton- polariton interactions are repulsive, it means that in the absence of an external magnetic field, the condensate is formed in a linear-polarized state, minimizing polariton-polariton repulsion \cite{Laussy2006,Kasprzak2007}. 

An important role of the polarization in the system of coupled condensates was already noted \cite{Ohadi2015,Pickup2021,Gnusov2021,Aristov2022}, but formulation of the problem in terms of extended version of XY model was never performed so far, up to our knowledge. Moreover, very recent experiments have demonstrated the ability to engineer beyond-nearest-neighbor couplings in spinor polariton condensate arrays \cite{PhysRevB.108.L161301}, further expanding the parameter space of accessible lattice Hamiltonians. In the current paper, we do the corresponding task, discussing the main differences of scalar and spinor models and considering important particular geometries of coupled spinor condensates.

\textit{Classical spin Hamiltonian.}
We consider the case of zero external magnetic field and start our consideration from the quantum Hamiltonian of the system of coupled condensates written in the secondary quantization representation:
\begin{multline}
\hat{H}=J\sum_{\langle jl\rangle,\sigma=\pm}a^\dagger_{j\sigma}a_{l\sigma}\\
+\delta J\sum_{\langle jl\rangle}\left[e^{i\theta_{lj}}a^\dagger_{j+}a_{l-}+e^{-i\theta_{lj}}a^\dagger_{l-}a_{j+}\right],
\end{multline}
where the operators $a_{j\sigma},a^\dagger_{j\sigma}$  correspond to polaritons with right and left circular poilarizations (referred to as spins, $\sigma=\pm$) located at site $j$of the lattice, $\langle jl\rangle$ indicates the summation by nearest neighbors only. 

In the above expression, the energy is counted from the energy of the creation of a single polariton, so the term corresponding to uncoupled condensates is omitted; the first term describes the spin-conserving tunneling between the neighboring lattice sites, and the second term refers to spin-flip tunneling. The latter is provided by TE-TM splitting, and the angles $\theta_{jl}$ are nothing but double angles between the x-axis and the line connecting the sites $j$ and $l$ \cite{Nalitov2015}. As we shall see later on, they have crucial effect on properties of the condensate. 

In our further consideration, we use the mean-field approach, in which the state of the coupled condensates is described by a direct product of macroscopically coherent wavefunctions corresponding to individual lattice sites,
\begin{equation}
|\Psi\rangle=\prod_{j,\sigma=\pm}|\alpha_{j\sigma}\rangle, \quad a_{j\sigma}|\alpha_{j\sigma}\rangle=\alpha_{j\sigma}|\alpha_{j\sigma}\rangle
\end{equation}
with order parameters being
\begin{equation}
\alpha_{j\sigma}=\sqrt{\frac{n}{2}}e^{i\varphi_{j\sigma}},
\end{equation}
where $n$ is the mean total occupancy of the condensate which is supposed to be the same for all condensates, and $\varphi_{j\sigma}$ are condensate phases. As we consider the case where magnetic field is absent, we suppose all condensates to be linearly polarized.  
Note that for a coherent state the vector of pseudospin is normalized to unity, $S_{jx}^2+S_{jy}^2+S_{jz}^2=1$.

The classical Hamiltonian of the problem can be computed as:
\begin{equation}\label{eq:classical_ham}
H(n_{j\sigma},\varphi_{j\sigma})=\langle\Psi|\hat{H}|\psi\rangle=H_1+H_2,
\end{equation}
where
\begin{eqnarray}
    H_1=\frac{Jn}{2}\sum_{\langle jl\rangle}[\cos(\varphi_{j+}-\varphi_{l+})+\cos(\varphi_{j-}-\varphi_{l-})]
\end{eqnarray}
is the part of the Hamiltonian, which describes the spin-dependent tunneling, and 
\begin{eqnarray}
    H_2=\frac{\delta J n}{2}\sum_{\langle jl\rangle}[\cos(\varphi_{j+}-\varphi_{l-}-\theta_{lj})+\\
    \nonumber+\cos(\varphi_{j-}-\varphi_{l+}+\theta_{lj})]
\end{eqnarray}
is the part of the Hamiltonian that stems from the spin-flip tunneling provided by the TE-TM splitting.

It is convenient to introduce the total and relative phases of the condensates as follows:
\begin{eqnarray}
   \Phi_j=\frac{\varphi_{j+}+\varphi_{j-}}{2},   \varphi_j=\varphi_{j+}-\varphi_{j-}.
\end{eqnarray}
Each of these two phases is equivalent to a two-dimensional spin located at the site $j$. The phase $\Phi_j$ corresponds to the spin $\bf{J}$ of an original scalar XY model,
\begin{equation}
\textbf{J}_j=\textbf{e}_x\cos \Phi_j+\textbf{e}_y\sin \Phi_j,
\end{equation}
while the phase $\phi_j$ defines the in-plane pseudospin components of the condensates (the z component is equal to zero),
\begin{equation}
\textbf{S}_j=\textbf{e}_x\cos \varphi_j+\textbf{e}_y\sin \varphi_j.
\end{equation}
The terms in the Hamiltonian can thus be rewritten as:
\begin{multline}
    H_1=Jn\sum_{\langle jl\rangle}\cos(\Phi_{j}-\Phi_{l})\cos(\frac{\varphi_{j}-\varphi_{l}}{2})\\
     =\frac{J_1}{\sqrt{2}}\sum_{\langle jl}(\textbf{J}_j\cdot \textbf{J}_l)\sqrt{1+(\textbf{S}_j\cdot \textbf{S}_l)},
\end{multline}
\begin{widetext} 
\begin{eqnarray}
    H_2=\delta Jn\sum_{\langle jl\rangle}\cos(\Phi_{j}-\Phi_{l})\cos(\frac{\varphi_{j}+\varphi_{l}}{2}-\theta_{ij})= \frac{J_2}{\sqrt{2}}\sum_{\langle jl\rangle}(\textbf{J}_j\cdot \textbf{J}_l)\sqrt{1+
    \cos(2\theta_{lj})(\textbf{S}_j\sigma_z\textbf{S}_l)+\sin(2\theta_{lj})(\textbf{S}_j\sigma_x\textbf{S}_l)},
\end{eqnarray}
\end{widetext}
where we introduced $J_1=Jn$, $J_2=\delta J n$, while $\sigma_z,\sigma_x$ are Pauli matrices. Note that the term $H_1$, corresponding to polarization-conservative tunneling, is rolled to zero when $\textbf{S}_j\cdot\textbf{S}_l=0$, i.e., the condensates are orthogonal polarized to each other, as expected. For further analysis, it will also be convenient for us to use the direction of polarization, the slope of which is determined by half the angle $\phi_j$ of each condensate.

\begin{figure}[t]
\includegraphics[width=1\columnwidth]{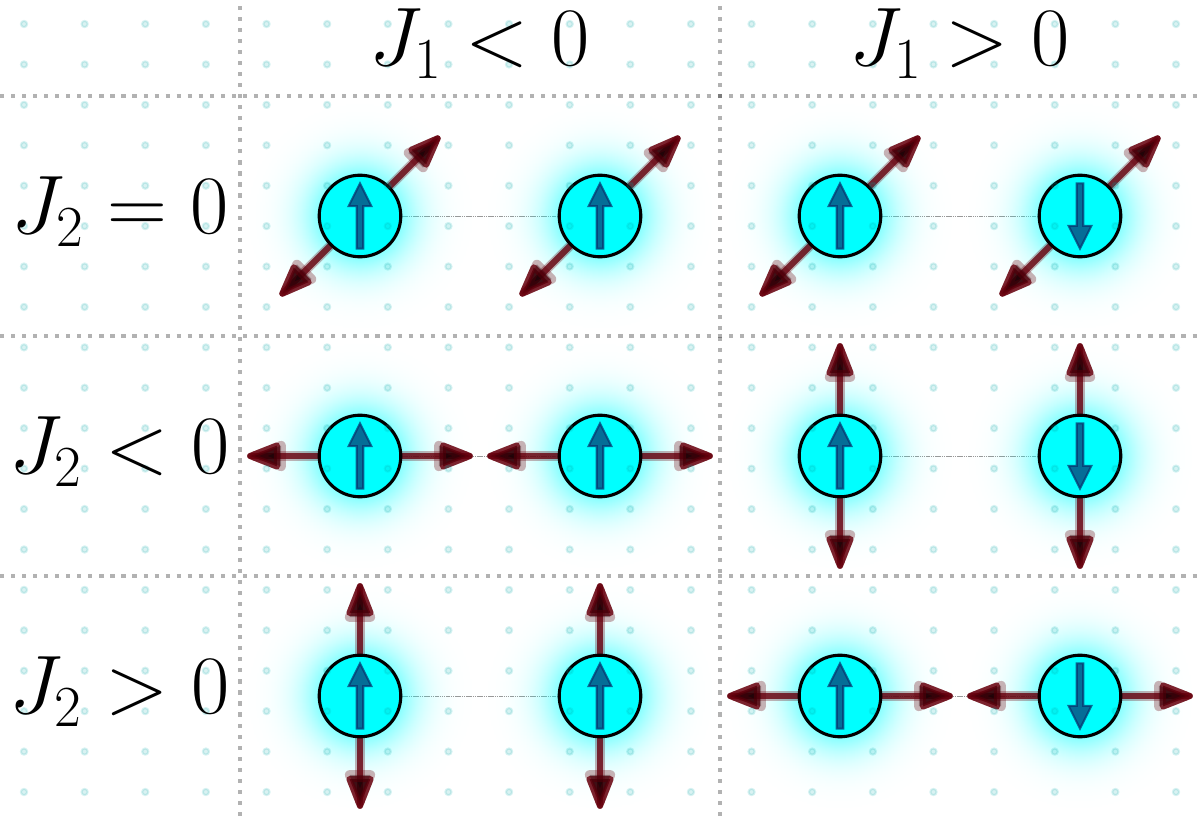}
\caption{Spin and polarization configurations for a two-site (dyad) system for various combinations of $J_1$ and $J_2$. Blue arrows inside the circles represent the $\mathbf{J}$-spins, while outer black double-headed arrows indicate the polarization directions. For $J_2=0$, the polarization is rotationally symmetric, but as soon as $J_2 \neq 0$, this symmetry is broken, and the polarization directions become pinned either along or perpendicular to the $x$-axis.
\label{fig:dyad}}
\end{figure}

\textit{Particular geometries}. We start from the simplest case of a dyad, i.e., two condensates coupled together, where we can put $\theta_{12}=0$. The latter condition means that the line connecting the two condensates is parallel to the $x$-axis.  The energy of the ground state can be described by the formula: $E=-2|J_1|-2|J_2|$. For ferromagnetic coupling ($J_1<0$) the total phases are the same, $\Phi_1=\Phi_2=\alpha$, where $\alpha$ is an arbitrary angle. This means that the spins $\textbf{J}_{1}$  and $\textbf{J}_{2}$ are parallel to each other as in the scalar case, but their orientation is arbitrary. As for the relative phases, they took the values $\varphi_{1}=\varphi_{2}=0$ for $J_2<0$ and $\varphi_{1}=\varphi_{2}=\pi$ for  $J_2>0$. This means that TE-TM splitting, resulting in the appearance of spin-flip tunneling terms, pins the direction of the liner polarization of the condensates parallel or perpendicular to the line connecting them. This is not surprising: In a dyad, the axes $X$ and $Y$ are no more equivalent, and in fact, polarization pinning should be expected \cite{Kasprzak2007}. This also happens for antiferromagnetic coupling ($J_1<0$), but now for the total phases we have $\Phi_1-\Phi_2=\pi$, which means that the spins $\textbf{J}_{1}$  and $\textbf{J}_{2}$ are anti-parallel (see Fig.~\ref{fig:dyad}). 

\begin{figure}[b]
\includegraphics[width=1\columnwidth]{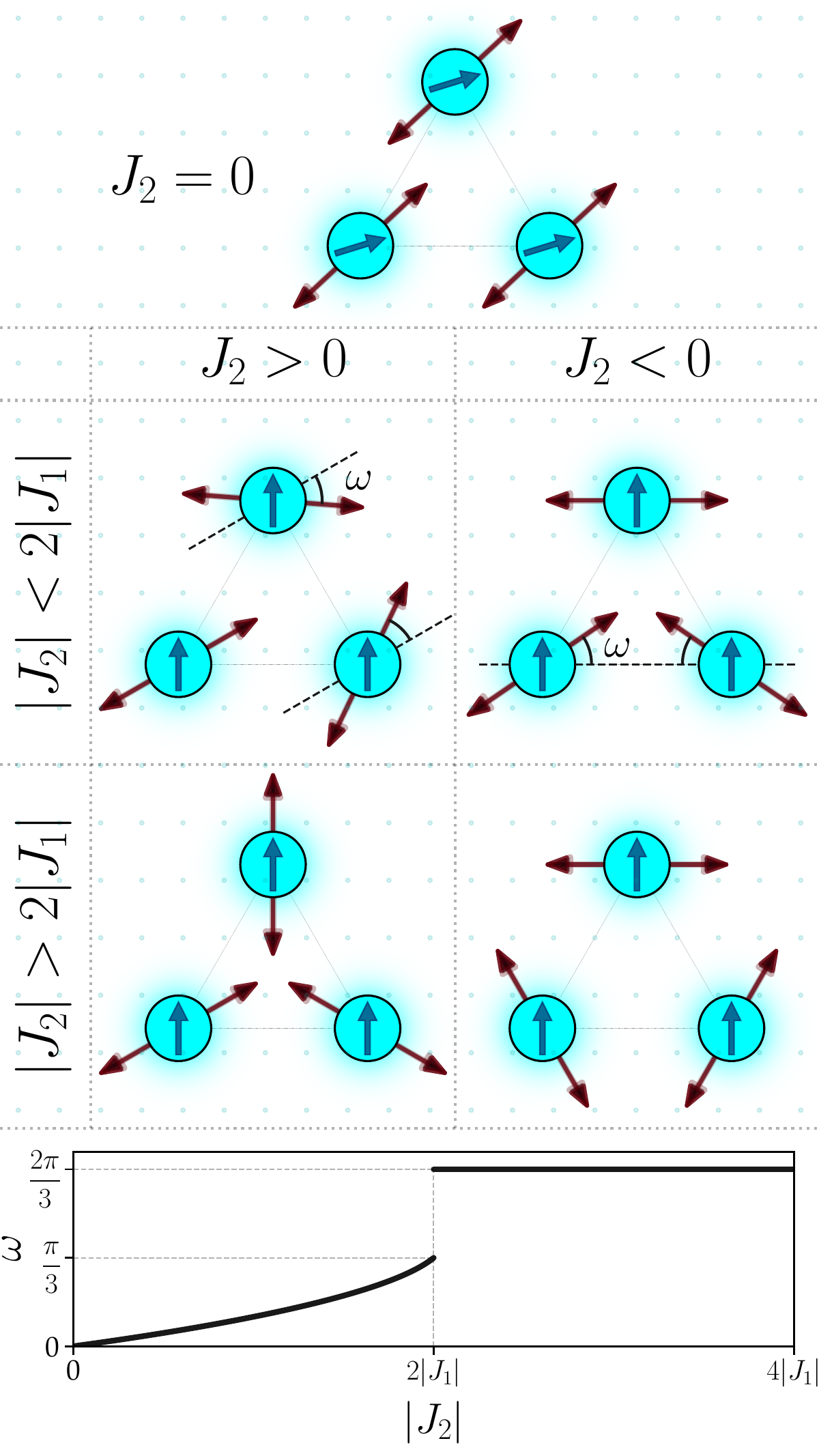}
\caption{Spin and polarization configurations for a triangle system for  $J_1<0$ (ferromagnetic coupling). For $J_2=0$, spins $\textbf{J}_l$ and polarizations of all the nodes are parallel, with total rotational symmetry present (upper panel). If $J_2\neq0$, the rotational symmetry for polarization is broken, polarization becomes pinned and polarization pattern tilted (middle two panels). Lower panel shows the dependence of the tilting angle $\omega$ on polarization-flip coupling amplitude $J_2$. Note, that after a gradual increase at $|J_2|=2J_1$, the tilting angle changes abruptly to $2\pi/3$, producing the patterns with polarizations aligned or perpendicular to the triangle bisectors for the cases of positive and negative $J_2$ respectively.
\label{fig:triangle_ferromagnetic}}
\end{figure}

Let us now consider the case of a triangle, where $\theta_{12}=2\pi/3$, $\theta_{13}=0$, and $\theta_{23}=4 \pi/3$. For the ferromagnetic case ($J_1<0$) possible spin and polarization configurations are presented in Fig.~\ref{fig:triangle_ferromagnetic}.  If $J_2=0$, we obtain the usual ferromagnetic spin alignment of ${\bf J}_i$ ($\Phi_i=\alpha$, for all $i=1,2,3$), and the polarizations of all condensates are parallel to each other and can be oriented in any direction. The non-vanishing value of $J_2$ does not affect the collinear ordering and rotational symmetry of spins ${\bf J}_i$, while it breaks the symmetry for the pseudospins ${\bf S}_i$, and preferential directions for the polarization appear. 

For infinitesimally small positive $J_2$ the energy is minimized in the configuration with $\varphi_1=\varphi_2=\varphi_3=(2n+1)\pi/3$ with $n=0,1,2$, which corresponds to the cases where the polarization is directed along one of the bisectors of the triangle. The increase of $J_2$ leads to the tilting of this simple polarization pattern so that the polarizations of the different nodes become non-collinear, as shown in the left-middle panel in Fig.~\ref{fig:triangle_ferromagnetic}.   Angles can be characterized as follows (we took the case corresponding to $n=0$, the cases of $n=1,2$ are similar): $(\phi_1,\phi_2,\phi_3)=(\pi/3,\pi/3-2\omega,\pi/3+2\omega)$, where the polarization tilt angle $\omega$ can be found by the following condition:
\begin{multline}
\dfrac{\sqrt{3}|J_2|}{J_1}\cos{\omega} + \left(\dfrac{|J_2|}{J_1}+2 + 4\cos{\omega}\right)\sin{\omega}=0,\label{eq:omega_condition}
\end{multline}

\noindent which was obtained from the minimization of Hamiltonian~\eqref{eq:classical_ham}. 

The dependence of $\omega$ on $J_2$ is shown in the lower panel of Fig.~\ref{fig:triangle_ferromagnetic}. The tilt angle first experiences a gradual increase, as expected, until it reaches the value $\omega=\pi/3$ at $J_2=2J_1$. At this point, it jumps to the value $\omega=\pi/3$, which corresponds to the orientation of the polarizations of all nodes along the corresponding triangle bisectors and does not change any further. 

\begin{figure}[t]
\includegraphics[width=1\columnwidth]{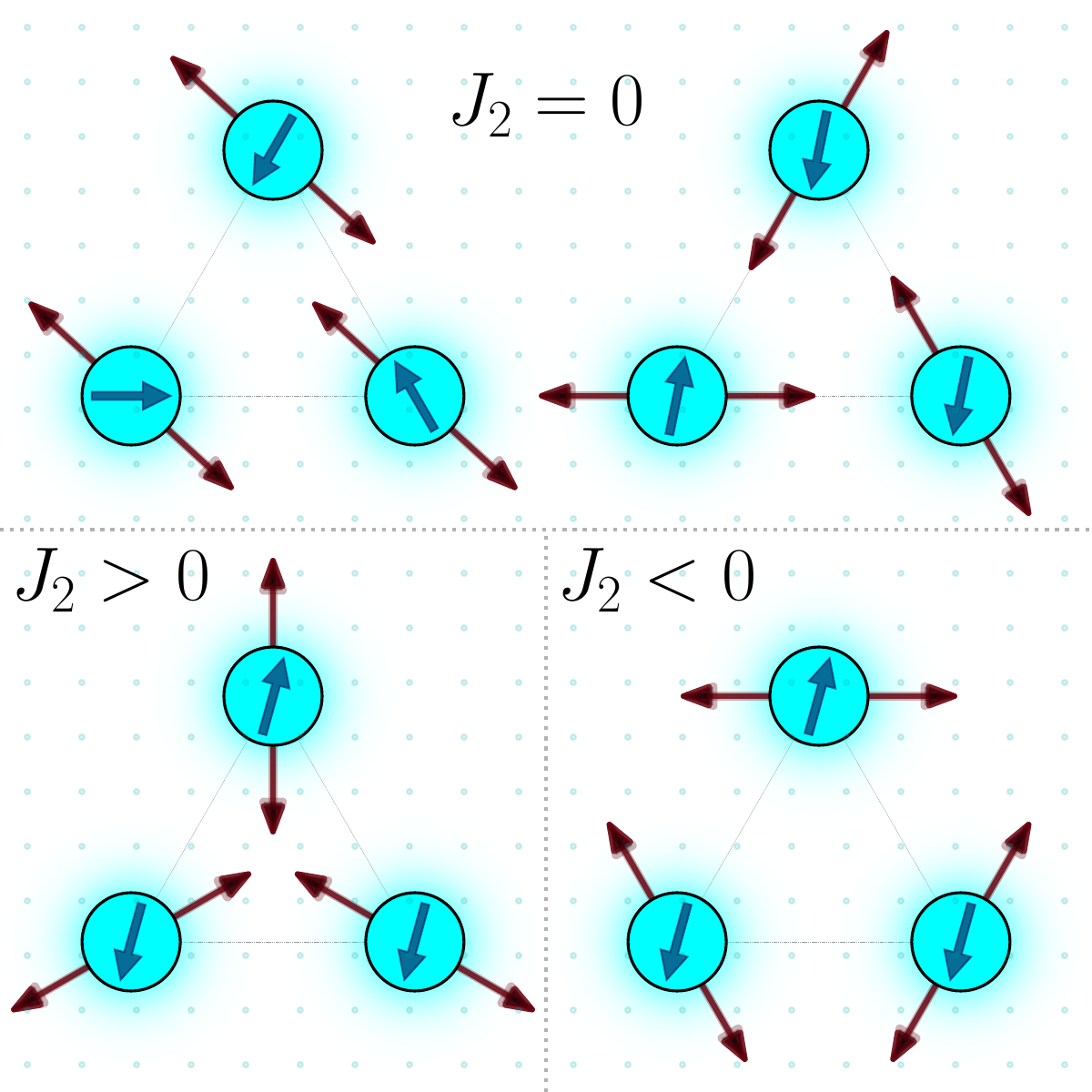}
\caption{
Spin and polarization configurations for a triangle system for  $J_1<0$ (ferromagnetic coupling). For $J_2=0$, two ground state configurations are possible: one with polarizations of all the nodes parallel tand spins $\textbf{J}_l$ making angles $2\pi/3$ with respect to each other, and another one, with two spins $\textbf{J}_l$ parallel, and third antiparallel, and polarizations making angle $\pi/3$ with respect to each other (upper panel). For finite values of $J_2$ only the latter configuration of spins $\textbf{J}_l$, and polarizations of the nodes become aligned with triangle bisectors or perpendicular to them depending on the sign of $J_2$ (lower panels). 
\label{fig:triangle_antiferromagnetic}}
\end{figure}
For infinitely small negative $J_2$ polarizations, all nodes are parallel, and all are oriented along one of the edges of the triangle, so that $\varphi_1=\varphi_2=\varphi_3=2n\pi/3$ with $n=0,1,2$. The increase in $|J_2|$ leads to the tilting of the pattern, so $(\phi_1,\phi_2,\phi_3)=(2\omega,0,-2\omega)$ for n = 0 (the cases of $n=1,2$ are fully equivalent). The tilt angle is defined by the same condition \ref{eq:omega_condition}, and its dependence on $|J_2|$ is identical to those presented for the case $J_2>0$ above. Note that for large negative values of $J_2$ polarizations of all nodes are perpendicular to the respective bisectors of the triangle.  

The dependence of the energy of the ground state on $|J_2|$ is identical for $J_2>0$ and $J_2<0$. Above the critical value $|J_2|=2J_1$, $\omega=2\pi/3$ it demonstrates the linear scaling and is described by simple analytical formula $E_3=3|J_1| - 6 |J_2|$. In the region $|J_2|<2J_1$ obtaining of such simple analytical expression is impossible, but following approximate universal formula describes the ground state energy with high accuracy:
\begin{align}
 E_3=-6 |J_1| - 0.542 \dfrac{J_2^2}{|J_1|} - 0.052 \dfrac{J_2^4}{|J_1^3|}.
\end{align}

Let us now consider the antiferromagnetic case ($J_1>0$). Here, even $J_2=0$, the situation here is more tricky than in the ferromagnetic case, as we face a situation of frustration.  For a scalar case, all spins $\mathbf{J}_l$ make angles $2\pi/3$ with each other. This configuration is also possible in the vector case considered here, provided that the polarizations of all nodes are parallel (see the upper left plot in Fig.~\ref{fig:triangle_antiferromagnetic}). However, other configurations with the same energy appear, those shown in the upper right panel of Fig.~\ref{fig:triangle_antiferromagnetic}). For them, the polarizations of the nodes make $\pi/3$ angles with each other, and the configuration of the spins $\mathbf{J}_l$ is as follows: two spins are parallel to each other, and the third one is antiparallel to both of them. Naturally, total rotational symmetry is present in the problem for $J_2=0$.

The finite value of $J_2$ splits the energies of these two possible configurations, and it is only the latter which corresponds to the ground state. This makes the usual scalar and vector case considered here qualitatively different. Naturally, breaking of the rotational symmetry also leads to a pinning of the polarization directions. For $J_2$ they are aligned with bisectors of the triangle, while for $J_2<0$ are perpendicular to them (see lower panels in Fig.~\ref{fig:triangle_antiferromagnetic}). The energy of the ground state demonstrates linear scaling with $|J_2|$, $E=-3J_1 - 6|J_2|$.

\begin{figure}[t]
\includegraphics[width=1\columnwidth]{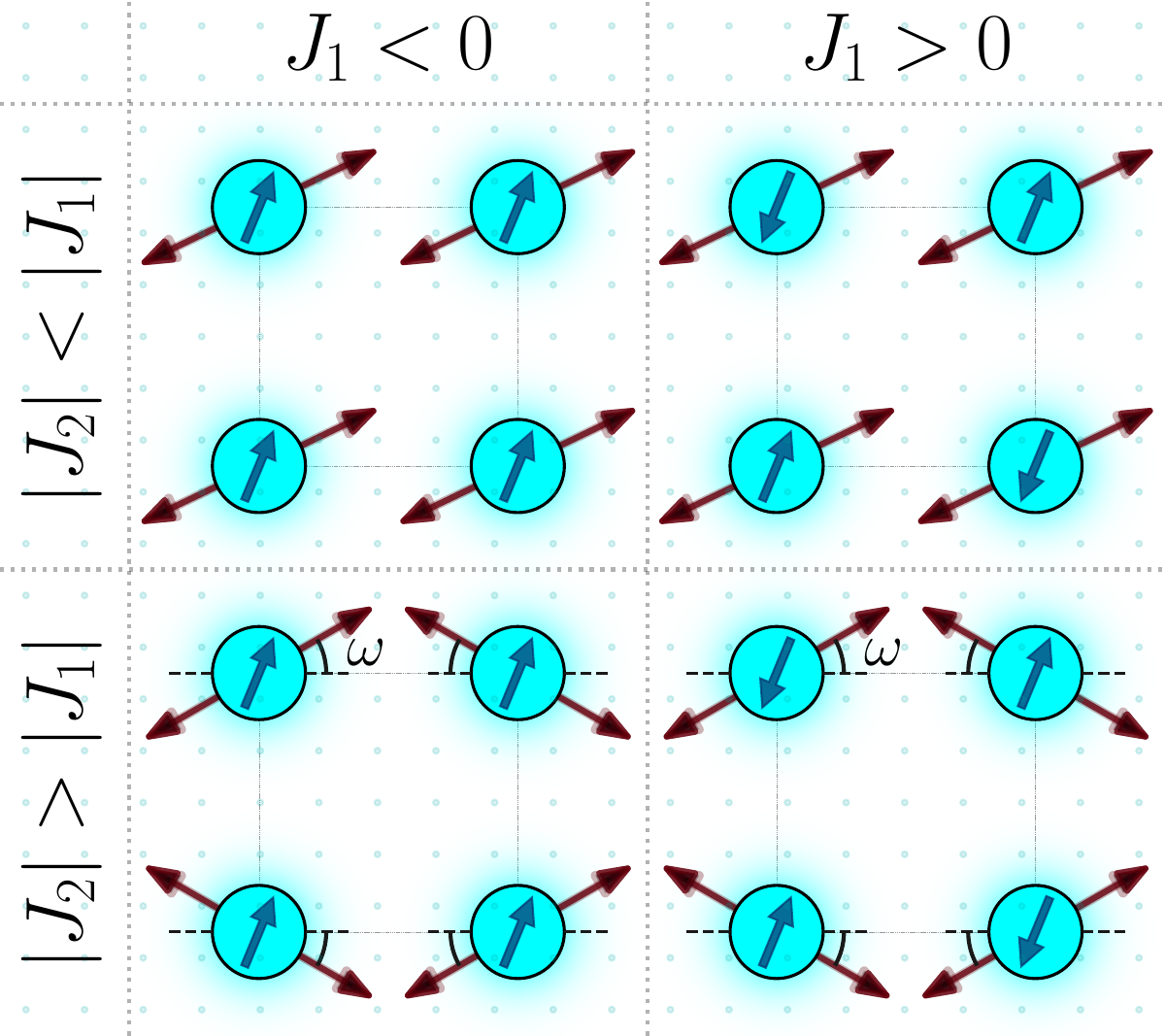}
\caption{Spin and polarization configurations for a square system for $J_1$ of both signs. Regardless of the value of $J_2$, the spins ${\bf J}_i$ are ordered collinear ($J_1<0$) and non-collinear ($J_1>0$) to each other and in an arbitrary direction in the plane. As for the polarizations, for $|J_2|<|J_1|$, they all parallel to each other and directed in arbitrary direction in plane  (upper   panel), while for $|J_2|>|J_1|$ a relationship arises between the directions of polarization, expressed through the angle $\omega$, measured from the $x$-axis (lower panel). 
\label{fig:4_nodes}}
\end{figure}

Finally, let us consider the case of a square. Unlike a triangle considered above, the geometry of the system does not support frustration, and thus the ferromagnetic and antiferromagnetic cases are qualitatively similar. For both of them, for $|J_2|<|J_1|$ polarizations are all collinear to each other, but their overall orientation is arbitrary. Spins $\mathbf{J}_l$ show the standard pattern of ferromagnetic or antiferromagnetic ordering, for negative and positive $J_1$, respectively.  For $|J_2|>|J_1|$ the polarizations are no longer collinear. They can be characterized by the following set of angles $(-\omega,\omega,-\omega,\omega)$ measured with respect to $x$-axis, while the pattern of the spins $\mathbf{J}_l$ remains unchanged. 

What makes the case of a square remarkable is the dependence of the energy of the ground state on $J_2$. The following universal formula can be obtained: 
\begin{equation}
 E=-8 \, {\rm max}(|J_1|,|J_2|)   
\end{equation}
This means that for $|J_2|<|J_1|$ the energy does not depends on $J_2$ at all, while for $|J_2|>|J_1|$ the linear scaling is recovered. The situation is similar to what is happening in spin Meissner effect, where the Zeeman splitting is fully screened by spin-anisotropic polariton-polariton interactions \cite{PLARubo}. In our case, the role of a real magnetic field is played by TE-TM splitting. which can be viewed as an effective magnetic field oriented in-plane, and the role of spin-anisotropic polariton-polarion interactions is played by tunneling term $H_1$, which is maximal for parallel polarization of the nodes, and is rolled to zero when they are orthogonally polarized.

\begin{figure}[b]
\includegraphics[width=1\columnwidth]{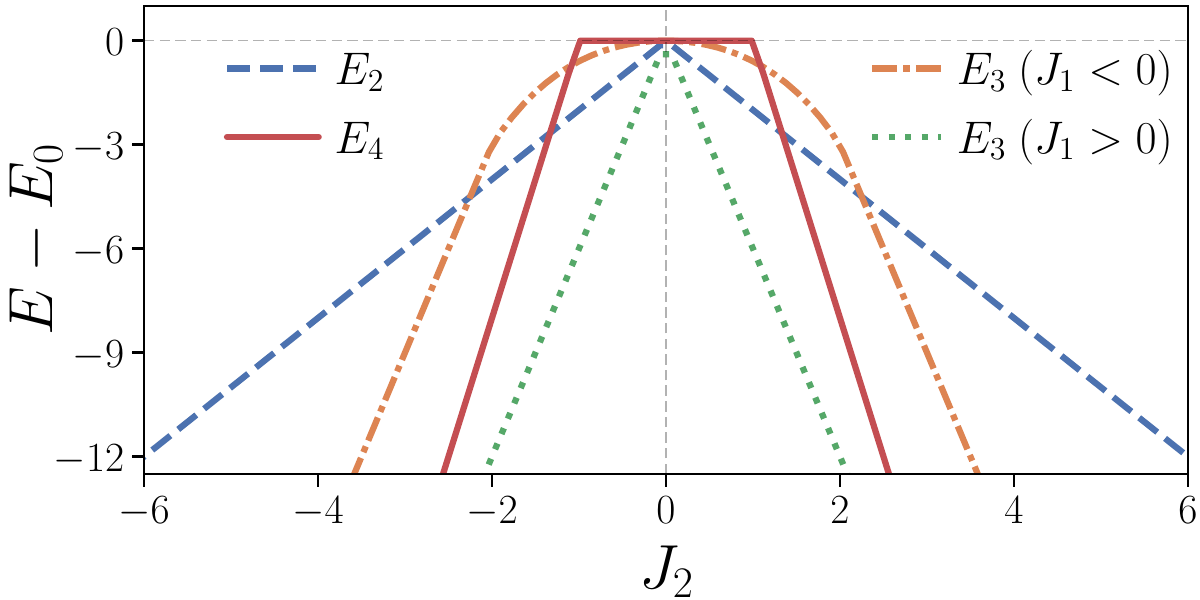}
\caption{Dependence of the ground-state energy $E$ on the coupling constant $J_2$ for different lattice configurations. The dashed line corresponds to a dyad, where the scaling is always linear. The dotted line corresponds to the triangle with $J_1>0$ (antiferromagnetic coupling), where linear scaling also occurs.  The dashed-dotted line illustrates the case of a triangle with ferromagnetic coupling $J_1<0$, where the dependence is quadratic at small $J_2$, but linear scaling is recovered for $|J_2|>2|J_1|$. Finally, the solid line gives the energy of the ground state of a square, where it does not depends on $J_2$ in the diapason $|J_2|<|J_1|$, and demonstrates linear scaling for $|J_2|>|J_1|$. This behaviou is similar to those happening in spin Meissner effect, where the role of a real magnetic field is played by TE-TM splitting. which can be viewed as an effective magnetic field oriented in-plane, and the role of spin-anisotropic polariton-polarion interactions is played by tunneling term $H_1$, which is maximal for parallel polarization of the nodes, and is rolled to zero when they are orthogonally polarized.
\label{fig:energies}}
\end{figure}

The dependencies of energies of a ground state on $J_2$ for three considered geometries are shown in Fig.~\ref{fig:energies}. As can be seen, this dependence is different for a dyad, a triangle, and a square, and only for the latter case is the energy independent of $J_2$ around $J_2=0$. This is a consequence of the particular symmetry of the TE-TM splitting.

\textit{Conclusions.}
We have developed and analyzed an extended XY model for tunnel-coupled spinor polariton condensates, accounting for both total phases (equivalent to 2D spins $\mathbf{J}$) and internal polarizations (equivalent to pseudospins $\mathbf{S}$). We demonstrated a crucial role for spin-flip tunneling terms originating from TE-TM splitting. In particular, we have shown that these terms can lead to reshaping of the structure of the ground state in triangle geometry. We also analyzed the dependence of the energy of the ground state on spin-flip tunneling amplitude and demonstrated that in the square geometry an analog of spin Meissner effect occurs, with TE-TM splitting playing the role of an effective magnetic field. 

\textit{Acknowledgements.}
A.K. acknowledges the support of the Icelandic Research Fund (Ranns\'oknasj\'o{\dh}ur, Grant No.~2410550). D.N. acknowledges the Ministry of Science and Higher Education of the Russian Federation (Goszadaniye) Project No. FSMG-2023-0011. 

\bibliography{references}   

\end{document}